\documentclass{article}%
\usepackage{algorithm}
\usepackage{algpseudocode}
\usepackage{alphabeta}
\usepackage{amsmath}
\usepackage{amsfonts}
\usepackage{amssymb}
\usepackage{boxedminipage}
\usepackage[open,openlevel=1]{bookmark}
\usepackage{caption}
\usepackage{color}
\usepackage{graphicx}
\usepackage{listings}
\usepackage[framed,numbered,autolinebreaks,useliterate]{mcode}
\usepackage[most]{tcolorbox}
\usepackage{multirow}
\usepackage{tikz}
\usepackage[framed,numbered,autolinebreaks,useliterate]{mcode}%
\providecommand{\U}[1]{\protect \rule{.1in}{.1in}}
\providecommand{\U}[1]{\protect \rule{.1in}{.1in}}
\numberwithin{equation}{section}
\newtheorem{theorem}{Theorem}[section]

\newtheorem{proposition}[theorem]{Proposition}

\begin{document}

\title{Markov Chains of Evolutionary Games \\with a Small Number of Players}
\author{Ath. Kehagias}
\date{\today}
\maketitle

\begin{abstract}
We construct and study the transition probability matrix of evolutionary games
in which the number of players is finite (and relatively small) of such games.
We use a simplified version of the population games studied by Sandholm. After
laying out a general framework we concentrate on specific examples, involving
the Iterated Prisoner's Dilemma, the Iterated Stag Hunt, and the
Rock-Paper-Scissors game. Also we consider several revision protocols: Best
Response, Pairwise Comparison, Pairwise Proportional Comparison etc. For each
of these we explicitly construct the MC transition probability matrix and
study its properties.

\end{abstract}

\section{Introduction\label{sec01}}

In this paper we are concerned with \emph{evolutionary games} in which the
number of players is finite (and relatively small). Our main goal is to
construct and study the \emph{transition probability matrix }of such games.
\emph{This is a preliminary exploration of the subject}; we will present more
extensive results in future publications.

Classical descriptions of evolutionary games appear in
\cite{Hofbauer1998,Sandholm2010a,Sandholm2010b}. Our main starting point is
\cite{Sandholm2010b}, but we use a simplified version of the model presented
therein. Also, we limit ourselves to games in which \emph{the number of
strategies available to the players is three}. Adhering to this (admittedly
limited) specification it is easy to see that every evolutionary game
generates a \emph{Markov chain} (MC)\ in which the \emph{state} is a vector
$\mathbf{s=}\left(  s_{1},s_{2},s_{3}\right)  $, where $s_{m}$ is the number
of players using the $m$-th strategy.

After laying out the general framework which describes our approach, we
concentrate on specific examples. We look at several possible \emph{base
games} such as \emph{Iterated Prisoner's Dilemma} (IPD), \emph{Iterated Stag
Hunt} (ISH), \emph{Rock-Paper-Scissors} (RPS); also we consider several
\emph{revision protocols} such \emph{Best Response} (BR), \emph{Pairwise
Comparison} (PC), \emph{Pairwise Proportional Comparison} (PPC) etc. For each
of these we explicitly construct the MC\ transition probability matrix and
study its properties.

\section{Evolutionary Games and Markov Chains\label{sec02}}

To construct an evolutionary game we start with a \emph{symmetric } two-player
game described by an $M\times M$ bimatrix $\left(  B,B^{T}\right)  $; i.e.,
each player can use a pure strategy from the set $\Sigma=\left \{  \sigma
_{1},...,\sigma_{M}\right \}  $ and we have%
\[
B_{m_{1}m_{2}}=\text{\textquotedblleft the payoff received by row player when
he uses }\sigma_{m_{1}}\text{ against column player using }\sigma_{m_{2}%
}\text{\textquotedblright.}%
\]

\noindent Now assume the existence of $N$ \emph{players} and define the following:

\begin{enumerate}
\item A \emph{population }vector $\mathbf{z}=\left(  z_{1},...,z_{N}\right)
\in \mathbf{Z}$ indicates the strategy used by each player, i.e., $z_{n}=m$
means the $n$-th player is using the $m$-th strategy; we will often call such
a player an \textquotedblleft$m$-\emph{user}\textquotedblright. It takes
values in the
\[
\text{Population Space}\text{: }\mathbf{Z}=\left \{  \left(  z_{1}%
,...,z_{N}\right)  :\forall n:z_{n}\in \left \{  1,..,M\right \}  \right \}  .
\]
We repeat that, in all cases studied in this paper we will have $M=3$ strategies.

\item A \emph{strategy }or \emph{state }vector $\mathbf{s}=\left(
s_{1},...,s_{M}\right)  \in \mathbf{S}$ indicates how many players are using
each strategy, i.e., $s_{m}$ is the number of $m$-users. It takes values in
the
\[
\text{State Space}\text{: }\mathbf{S}=\left \{  \left(  s_{1},...,s_{M}\right)
\in \left \{  0,1,..,N\right \}  \text{ such that }\sum_{m=1}^{M}s_{m}=N\right \}
.
\]

\end{enumerate}

\noindent For given payoff matrix $B$, and number of players $N$, the
evolutionary game consists of the following procedure.

\bigskip

\noindent \fbox{\begin{minipage}{0.98\textwidth}
\parbox{0.97\textwidth}{
\medskip
\begin{center}
\underline{\textbf{Evolutionary Game Protocol}}
\end{center}
\begin{enumerate}
\item Choose an initial population vector $\mathbf{z}\left(
0\right)  =\left(  z_{1}\left(  0\right)  ,...,z_{N}\left(  0\right)  \right)
\in \mathbf{Z}$, i.e., an initial distribution of strategies to players.
\item Repeat for $j\in \left \{  0,1,...,J-1\right \}  $ the following iteration.
\begin{enumerate}
\item Compute the state vector $\mathbf{s}\left(  j\right)  =\left(
s_{1}\left(  j\right)  ,...,s_{M}\left(  j\right)  \right)  \in \mathbf{S}$ and
the \emph{frequency vector} $\mathbf{x}\left(  j\right)  =\left(  x_{1}\left(
j\right)  ,...,x_{M}\left(  j\right)  \right)  $ as follows:\[
\forall m:s_{m}\left(  j\right)  =\left \vert \left \{  z_{n}\left(  j\right)
:z_{n}\left(  j\right)  =m\right \}  \right \vert ,\qquad x_{m}\left(  j\right)
=\frac{s_{m}\left(  j\right)  }{\sum_{m=1}^{M}s_{m}\left(  j\right)  }.
\]
\item The players play all possible matches between each other, i.e., for all
the pairs $\left(  n_{1},n_{2}\right)  $ with $n_{1},n_{2}\in \left \{
1,...,N\right \}  $ and $n_{1}\neq n_{2}$. In each match they play  one round of the $B$ game.
\item The $n$-th player (for $n\in \left \{  1,...,N\right \}  $) collects a
total payoff
\[
q_{n}\left(  j\right)  =\sum_{n\neq n^{\prime}}B_{z_{n}\left(  j\right)
z_{n^{\prime}\left(  j\right)  }},
\]
i.e., the $n$-th player collects his total payoff by playing strategy $z_{n}$
against the $n^{\prime}$-th player's strategy $z_{n^{\prime}}$, for every
$n^{\prime}\neq n$.
\item The $m$-th strategy collects a total payoff
\[
Q_{m}\left(  j\right)  =\sum_{n:z_{n}=m}q_{n}\left(  j\right)  ,
\]
i.e., the total payoff of the $m$-th strategy is the sum of the payoffs of the
$m$-users.
\item The next \ population vector $\mathbf{z}\left(  j+1\right)  $
(also called a \emph{generation})\ is obtained by letting some players update their
strategy. This is achieved by a \emph{revision protocol} $\mathcal{R}$; we
will presently discuss several such protocols.
\end{enumerate}
\end{enumerate}
}
\end{minipage}}

\bigskip

\noindent So an evolutionary game is a pair $\left(  B,\mathcal{R}\right)  $
(with auxiliary parameters $N$ and $J$) and produces sequences $\left(
\mathbf{z}\left(  j\right)  \right)  _{j=0}^{\infty}$, $\left(  \mathbf{s}%
\left(  j\right)  \right)  _{j=0}^{\infty}$ and $\left(  \mathbf{x}\left(
j\right)  \right)  _{j=0}^{\infty}$. As will be seen presently, for the
revision protocols used in the current paper, both the population process
$\left(  \mathbf{z}\left(  j\right)  \right)  _{j=0}^{\infty}$ and the state
process $\left(  \mathbf{s}\left(  j\right)  \right)  _{j=0}^{\infty}$ are
\emph{Markov chains} (MC); the same is true of the \emph{frequency process
}$\left(  \mathbf{x}\left(  j\right)  \right)  _{j=0}^{\infty}$ (which is
simply a function of $\left(  \mathbf{s}\left(  j\right)  \right)
_{j=0}^{\infty}$). By choosing revision protocols which favor better
performing strategies (i.e., those with higher total payoff), we hope that
$\left(  \mathbf{s}\left(  j\right)  \right)  _{j=0}^{\infty}$ will converge
into a state $\overline{\mathbf{s}}$ in which all players are using the best
performing strategy (more accurately: \emph{one} the best performing strategies).

Before proceeding, let us introduce some notations. We write the player total
payoff vector as $\mathbf{q}=\left(  q_{1},...,q_{N}\right)  $ and the
strategy total payoff vector as $\mathbf{Q}=\left(  Q_{1},...,Q_{M}\right)  $.
Note that both $\mathbf{q}$ and $\mathbf{Q}$ are functions of the state vector
$\mathbf{s}$ (or, equivalently, the frequency vector $\mathbf{x}$), the payoff
matrix $B$ etc. However, for the sake of brevity, we will usually omit this
dependence from our notation. We let $\overline{Q}$ be the average strategy
payoff and $\widehat{Q}$ be the maximum strategy payoff%
\[
\overline{Q}=\sum_{m=1}^{M}x_{m}Q_{m},\quad \widehat{Q}=\max_{m}Q_{m}.
\]
Also we use the notation $\left[  y\right]  _{+}=\max \left(  y,0\right)  $,
i.e., $\left[  y\right]  _{+}$ is the nonnegative part of $y$. Finally,
$\mathbf{1}_{\mathbf{U}}\left(  u\right)  $ is the indicator function of set
$\mathbf{U}$, i.e., it equals $1$ when the $u\in \mathbf{U}$ and $0$ when
$u\not \in \mathbf{U}$.

Let us now return to the revision protocols. All protocols used in this paper
follow the same general pattern. First, a player is selected equiprobably from
the population. Then, if the player is an $m_{1}$-user, he adopts the $m_{2}%
$-th strategy with probability $R_{m_{1}m_{2}}\left(  \mathbf{s}\right)  $,
where $R\left(  \mathbf{s}\right)  $ is an $M\times M$ \emph{rate matrix}
depending on the current state $\mathbf{s}$. Hence our revision protocols are
distinguished by the rate matrix; a particular $R\left(  \mathbf{s}\right)  $
determines a particular revision protocol. Here are some rate matrices.

\begin{enumerate}
\item \textbf{Best Response (BR)}. Strategy $m_{1}$ switches equiprobably to
the strategy of one of the best performing players. I.e., letting%
\begin{align*}
\text{the set of the players who achieve max payoff}  &  \text{: }%
\widehat{\mathbf{N}}=\left \{  n:q_{n}=\max_{i}q_{i}\right \}  ,\\
\text{the set of the strategies of these players}  &  \text{: }\widehat
{\mathbf{M}}=\left \{  z_{n}:n\in \widehat{\mathbf{N}}\right \}  ,
\end{align*}
we set
\[
R_{m_{1}m_{2}}=\frac{\mathbf{1}_{\widehat{\mathbf{M}}}\left(  m_{2}\right)
}{\left \vert \widehat{\mathbf{M}}\right \vert }.
\]

\item \textbf{Pairwise Proportional Comparison (PPC)}. Strategy $m_{1}$
switches to $m_{2}$ with probability proportional to the positive part of the
surplus of $Q_{m_{2}}$ over $Q_{m_{1}}$, multiplied by the frequency of
$m_{2}$:%
\[
R_{m_{1}m_{2}}=\frac{x_{m_{2}}\left[  Q_{m_{2}}-Q_{m_{1}}\right]  _{+}}%
{\sum_{m=1}^{M}x_{m}\left[  Q_{m}-Q_{m_{1}}\right]  _{+}}.
\]

\item \textbf{Pairwise Comparison (PC)}. Strategy $m_{1}$ switches to $m_{2}$
with probability proportional to the positive part of the surplus of
$Q_{m_{2}}$ over $Q_{m_{1}}$:%
\[
R_{m_{1}m_{2}}=\frac{\left[  Q_{m_{2}}-Q_{m_{1}}\right]  _{+}}{\sum_{m=1}%
^{M}\left[  Q_{m}-Q_{m_{1}}\right]  _{+}}.
\]

\item \textbf{Comparison to the Average Payoff (CAP)}. Strategy $m_{1}$
switches to $m_{2}$ with probability proportional to positive part of the
$Q_{m_{2}}$ surplus over $\overline{Q}$:%
\[
R_{m_{1}m_{2}}=\frac{\left[  Q_{m_{2}}-\overline{Q}\right]  _{+}}{\sum
_{m=1}^{M}\left[  Q_{m}-\overline{Q}\right]  _{+}}.
\]

\item \textbf{Logit}. Strategy $m_{1}$ switches to $m_{2}$ with the following
probability:%
\[
R_{m_{1}m_{2}}=\frac{\exp \left(  Q_{m_{2}}/\eta \right)  }{\sum_{m=1}^{M}%
\exp \left(  Q_{m}/\eta \right)  }%
\]
where $\eta$ is a noise parameter (lower $\eta$ values lead to best performing
strategies being selected with higher probability; higher $\eta$ values lead
to all strategies being selected with more or less equal proabbility).
\end{enumerate}

\bigskip

\noindent It is worth pointing out that all of the above rate matrices assign
zero probability to the adoption of a strategy which is not represented in the
population (why?). In other words, once a strategy becomes extinct in an
evolutionary game, it will never reappear. Having selected a rate matrix $R$,
we can compute the transition probability matrix $P$ of the $\left(
\mathbf{s}\left(  j\right)  \right)  _{j=0}^{\infty}$ process as follows%
\[
\forall \mathbf{s},\mathbf{s}^{\prime}:P_{\mathbf{ss}^{\prime}}=\Pr \left(
\mathbf{s}\left(  j+1\right)  =\mathbf{s}^{\prime}|\mathbf{s}\left(  j\right)
=\mathbf{s}\right)  =\left \{
\begin{array}
[c]{ll}%
\frac{s_{m_{1}}}{\sum_{m=1}^{M}s_{m}}R_{m_{1}m_{2}}\left(  \mathbf{s}\right)
& \text{if }s_{m_{1}}^{\prime}=s_{m_{1}}-1\text{ and }s_{m_{2}}^{\prime
}=s_{m_{2}}+1\\
0 & \text{else}%
\end{array}
\right.
\]
Actually, the above formula holds only for \textquotedblleft
interior\textquotedblright \ states, i.e., those with $s_{m}\in \left \{
2,...,N-1\right \}  $ for all $m$. The modifications for \textquotedblleft
boundary\textquotedblright \ states are straightforward; for example, with
$N=3$: the only possible transition from $\left(  3,0,0\right)  $ is to
itself, the possible transitions from $\left(  2,1,0\right)  $ are to $\left(
3,0,0\right)  $ and to $\left(  1,2,0\right)  $ etc.

\section{Prisoner's Dilemma\label{sec03}}

\subsection{The Base Game\label{sec0301}}

We are particularly interested in $B$ games obtained from an iterated
two-player \emph{base game}, described by a bimatrix $A$. From such an $A$ we
construct an $M\times M$ matrix $B$ where
\[
B_{m_{1}m_{2}}=\text{\textquotedblleft the payoff received by playing }T\text{
rounds of }A\text{, using strategy }\sigma_{m_{1}}\text{ against strategy
}\sigma_{m_{2}}\text{\textquotedblright.}%
\]
In this example we will suppose that $A$ is the Prisoner's Dilemma
\cite{Axelrod1981} with bimatrix
\[
\left(  A,A^{T}\right)  =\left(  \left[
\begin{array}
[c]{cc}%
3 & 1\\
4 & 2
\end{array}
\right]  ,\left[
\begin{array}
[c]{cc}%
3 & 4\\
1 & 2
\end{array}
\right]  \right)  .
\]
As mentioned in the Introduction, we will only use three strategies, and this
case they are:\ 

\begin{enumerate}
\item $\sigma_{1}=$\textquotedblleft AllC\textquotedblright:\ cooperate (i.e.,
play row or column one) in every round.

\item $\sigma_{2}=$\textquotedblleft AllD\textquotedblright:\ defect (i.e.,
play row or column two) in every round.

\item $\sigma_{3}=$\textquotedblleft TitForTat\textquotedblright: cooperate in
the first round; in every other round play what \ the other player played in
the previous round.
\end{enumerate}

\noindent This is a reasonable choice of strategies: we have a totally
cooperating strategy, a totally defecting one and a strategy which rewards
cooperation and punishes defection. \footnote{We could easily have used other
strategy sets. All experiments presented in the current paper have been
implemented with our \emph{Evolutionary Games Toolkit} (\textsf{EGT}) which
can be found at
\texttt{https://github.com/thanasiskehagias/EvolutionaryGamesToolkit}. It
contains many additional strategies and functionalities.}

Further, we assume that the iterated game is played for $T=1000$ rounds.
Consider $B_{11}$, the payoff of a player who uses \textquotedblleft
AllC\textquotedblright \ for 1000 rounds, against another player who also uses
\textquotedblleft AllC\textquotedblright; hence the first player will receive
\[
B_{11}=1000\cdot3=3000.
\]
similarly, $B_{12}$ is the payoff of a player who uses \textquotedblleft
AllC\textquotedblright \ for 1000 rounds, against another player who uses
\textquotedblleft AllD\textquotedblright; hence the first player will receive
\[
B_{11}=1000\cdot1=1000.
\]
\ Continuing in this manner we obtain%
\[
B=\left[
\begin{array}
[c]{ccc}%
3000 & 1000 & 3000\\
4000 & 2000 & 2002\\
3000 & 1999 & 3000
\end{array}
\right]
\]
and we have obtained our base game.

\subsection{Best Response Protocol\label{sec0302}}

We now obtain a specific evolutionary game by using $B$ in conjuction with the
BR revision protocol; furthermore, we take the number of players to be $N=3$.
It is then easy to compute the corresponding transition probability matrix
$P$; from this we can obtain the \emph{state transition garph}, illustrated in
Figure \ref{fig001}.a. In Figure \ref{fig001}.b we plot a realization of the
strategy process $\mathbf{s}\left(  1\right)  ,...,\mathbf{s}\left(
100\right)  $. \noindent To interpret the STG plot, keep in mind the following.

\begin{enumerate}
\item Every state $\mathbf{s}=\left(  s_{1},s_{2},s_{3}\right)  $ corresponds
to a dot of the diagram; the coordinates of the dot are $\left(  s_{1}%
,s_{2}\right)  $.

\item The diagram shows three absorbing states, the ones which have no
outgoing arrows. They have coordinates $\left(  0,0\right)  $, $\left(
3,0\right)  $ and $\left(  0,3\right)  $ so they correspond to the ``pure''
states $\left(  0,0,3\right)  $, $\left(  3,0,0\right)  $ and $\left(
0,3,0\right)  $. No other absorbing states exist, so we know from Markov chain
theory \cite{Kemeny1966} that, with probability one, the population will
eventually settle to one of the pure states.

\item Consequently, with every state $\mathbf{s}=\left(  s_{1},s_{2}%
,s_{3}\right)  $ of the STG\ is associated a vector $\left(  p_{1}\left(
\mathbf{s}\right)  ,p_{2}\left(  \mathbf{s}\right)  ,p_{3}\left(
\mathbf{s}\right)  \right)  $, where $p_{1}\left(  \mathbf{s}\right)  $ is the
probability that the process, when starting at $\mathbf{s}$, will be absorbed
by $\left(  3,0,0\right)  $; similarly $p_{2}\left(  \mathbf{s}\right)  $ is
the probability of absorption by $\left(  0,3,0\right)  $ and $p_{3}\left(
\mathbf{s}\right)  $ is the probability of absorption by $\left(
0,0,3\right)  $.

\item The dot of the STG\ associated with state $\mathbf{s}$ is color coded as
follows:\ we take $\left(  p_{1}\left(  \mathbf{s}\right)  ,p_{2}\left(
\mathbf{s}\right)  ,p_{3}\left(  \mathbf{s}\right)  \right)  $ to be an RGB
color vector and color the corresponding dot accordingly. For example $\left(
3,0,0\right)  $ is absorbed into itself with probability one, hence it is
painted red; similarly $\left(  0,3,0\right)  $ is painted green and $\left(
0,0,3\right)  $ blue. The color of dots close to $\left(  0,0\right)  $
(corresponding to states close to $\left(  0,0,3\right)  $)\ is close to blue,
because the probability of being absorbed by $\left(  0,0,3\right)  $ is equal
to or very close to one, as seen by the STG\ arrows. Similarly dots close to
$\left(  3,0\right)  $ (corresponding to states close to $\left(
3,0,0\right)  $)\ are close to red, because the probability of being absorbed
by $\left(  3,0,0\right)  $ is equal to or very close to one. On the other
hand, dot $\left(  5,5\right)  $ (i.e., state $\left(  5,5,5\right)  $) is
colored \textquotedblleft green-blue\textquotedblright, because (as seen by
following the STG\ arrows)\ has about equal probability of being absorbed by
either $\left(  0,0,3\right)  $ or $\left(  0,3,0\right)  $.
\end{enumerate}

\noindent

\begin{minipage}{1.0\textwidth}
\centering
\includegraphics[scale=0.45]{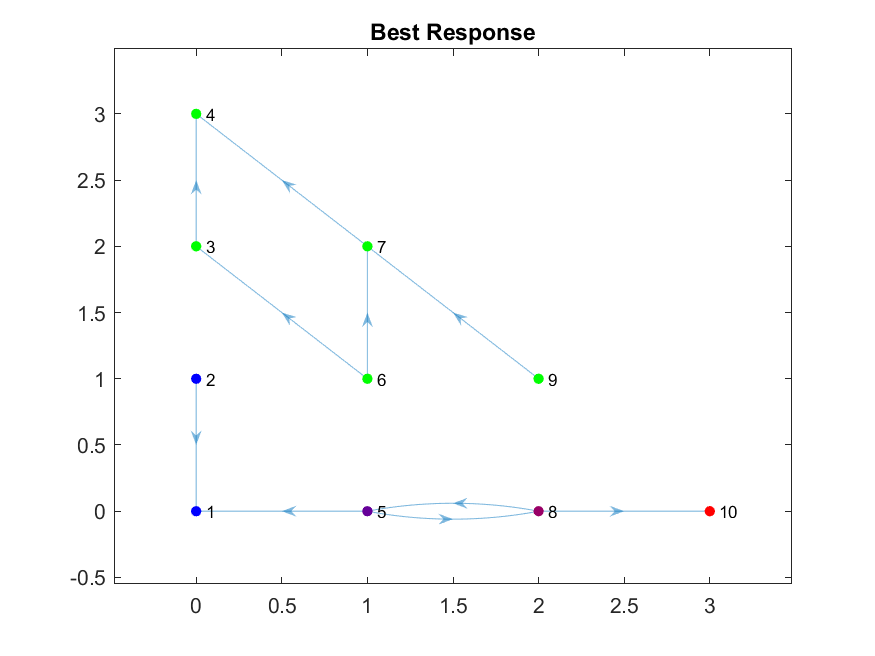}
\includegraphics[scale=0.45]{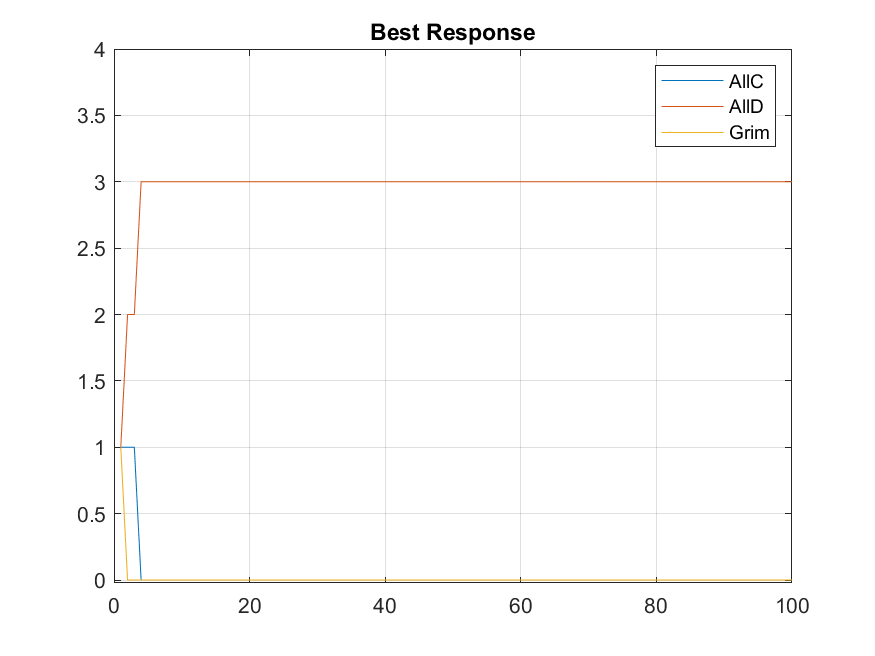}
\captionof{figure}{
Markovian analysis of an IPD evolutionary game with $N=3$ players. Left: STG of the Markov chain.
Right: plots of the numbers of $m$-users (for $m\in \{1,2,3\}$)
as a function of generation number; this is a realization of the Markov chain defined by
the evolutionary game.
}
\label{fig001}%
\end{minipage}

\vspace{5mm}

\noindent We repeat the experiment using $N=15$ players and we get the plots
of Figure \ref{fig002}.

\bigskip

\begin{minipage}{1.0\textwidth}
\centering
\includegraphics[scale=0.45]{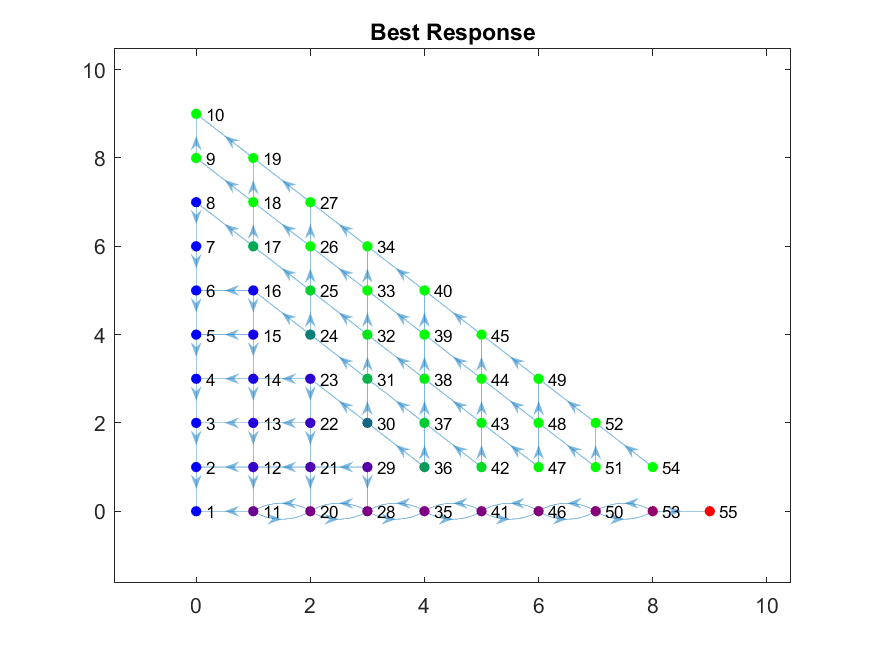}
\includegraphics[scale=0.45]{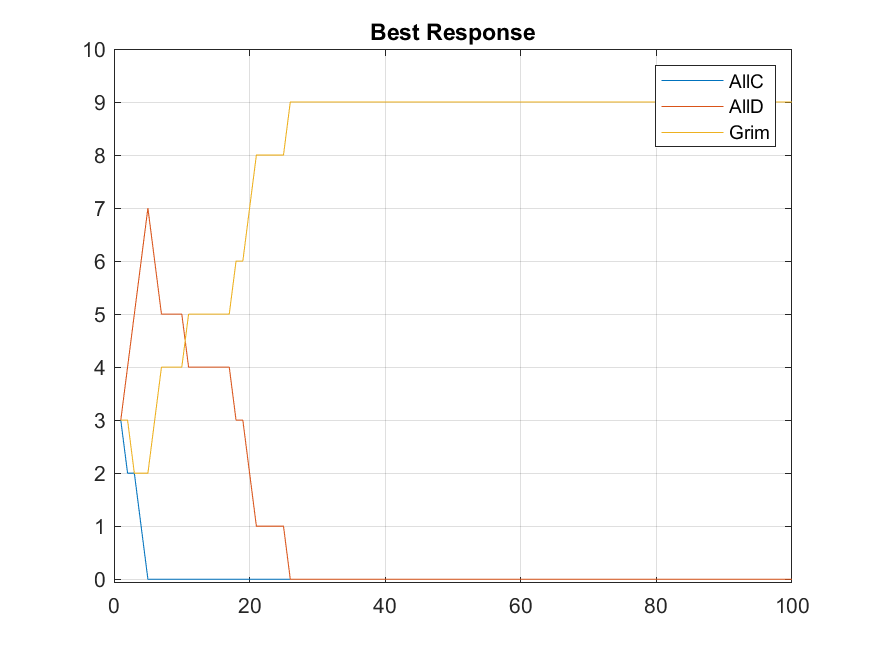}
\captionof{figure}{
Markovian analysis of an IPD evolutionary game with $N=9$ players. Left: STG of the Markov chain.
Right: plots of the numbers of $m$-users (for $m\in \{1,2,3\}$)
as a function of generation number; this is a realization of the Markov chain defined by
the evolutionary game.
}
\label{fig002}%
\end{minipage}

\vspace{5mm}

\noindent Finally, we repeat the experiment using $N\in \{12, 15, 30, 60\}$
players and we get the STG plots of Figure \ref{fig003}.

\vspace{5mm}

\begin{minipage}{1.0\textwidth}
\centering
\includegraphics[scale=0.45]{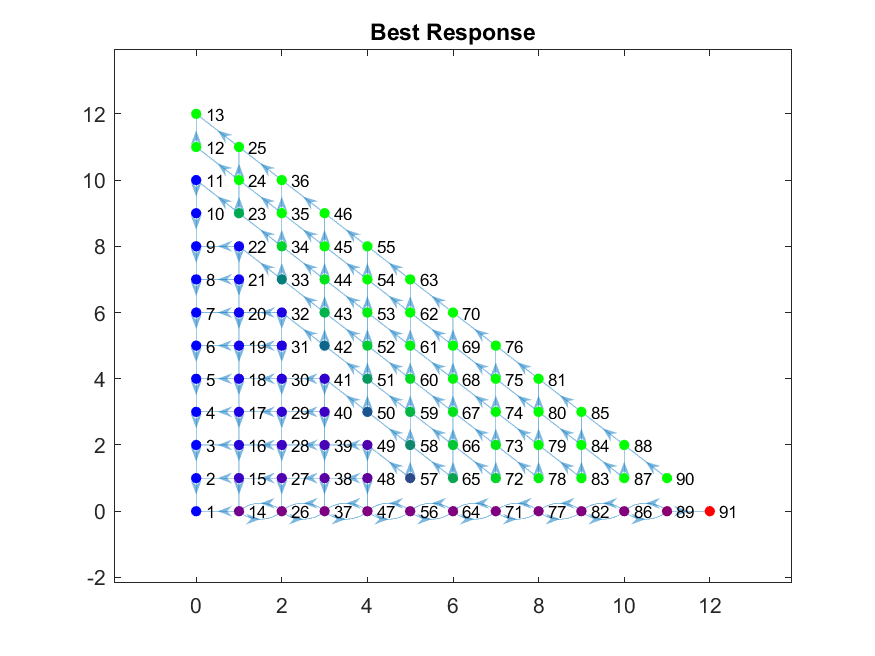}
\includegraphics[scale=0.45]{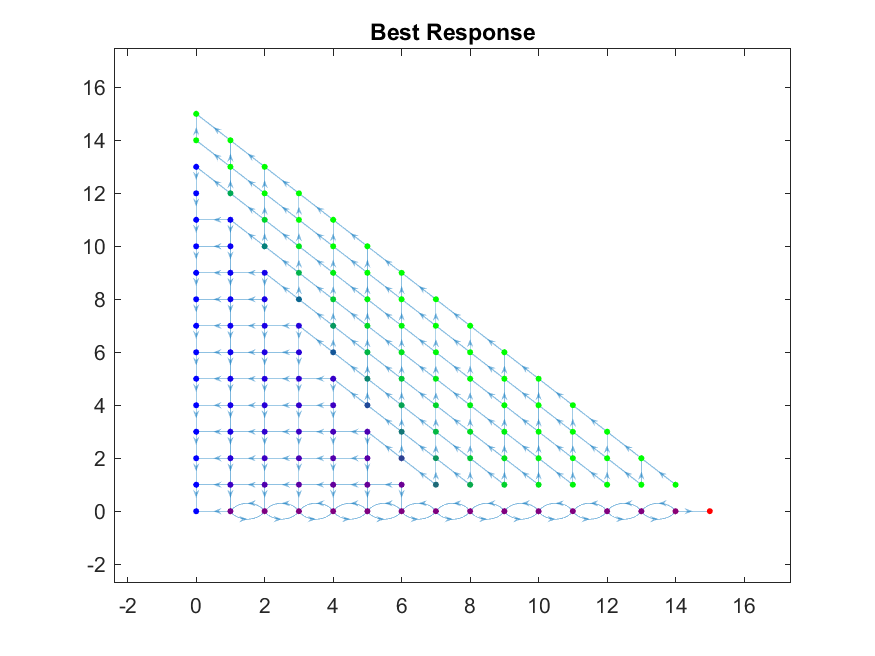}
\includegraphics[scale=0.45]{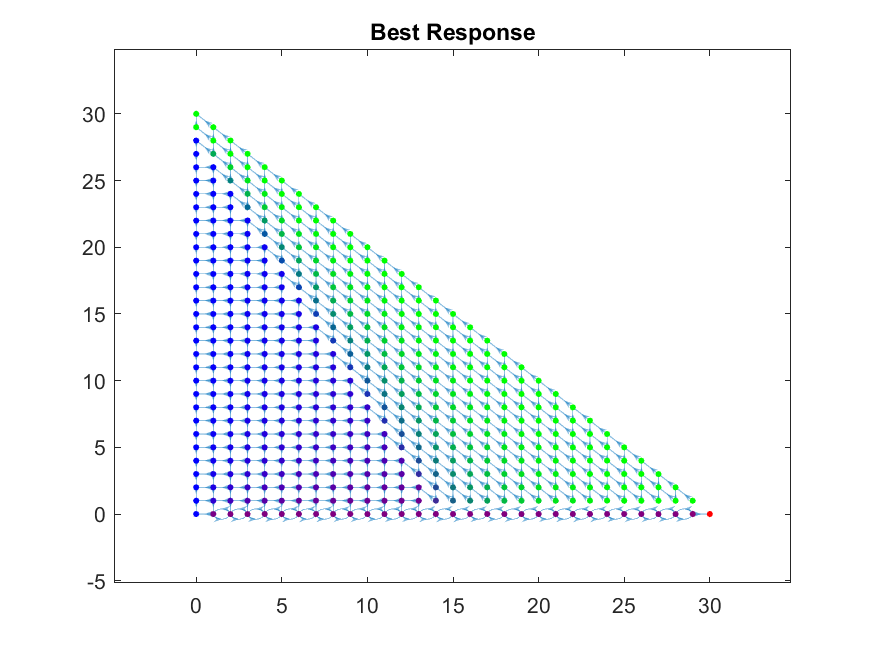}
\includegraphics[scale=0.45]{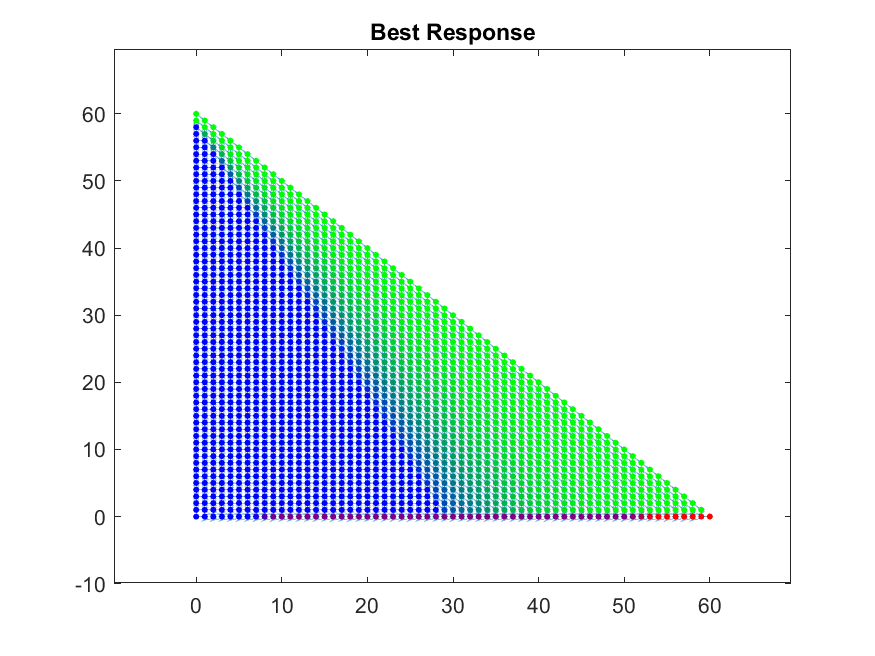}
\captionof{figure}{
STG diagrams for $N\in \{12,15,30,60\}$ players.
}
\label{fig003}%
\end{minipage}

\vspace{5mm}

\noindent The above plots suggest the following propositions; the proofs will
be provided in a future publication.

\begin{proposition}
\normalfont In the evolutionary game with IPD\ with $T$ rounds as base game,
best response revision protocol, and $N$ players, let $\widehat{\sigma
}^{\left[  s_{1},s_{2},s_{3}\right]  }$ denote the best strategy at state
$\left(  s_{1},s_{2},s_{3}\right)  $. Then, for all $T>2N$ and for all
$\left(  s_{1},s_{2},s_{3}\right)  \in S$ we have

\begin{enumerate}
\item When $s_{2}>0$:
\begin{align}
N-1  &  \leq2s_{1}+s_{2}\Rightarrow \widehat{\sigma}^{\left[  s_{1},s_{2}%
,s_{3}\right]  }=2,\label{eq04021}\\
N-1  &  >2s_{1}+s_{2}\Rightarrow \widehat{\sigma}^{\left[  s_{1},s_{2}%
,s_{3}\right]  }=3. \label{eq04022}%
\end{align}

\item When $s_{2}=0$:%
\[
\widehat{\sigma}^{\left[  s_{1},s_{2},s_{3}\right]  }\subseteq \left \{
1,3\right \}  .
\]

\end{enumerate}
\end{proposition}

\begin{proposition}
\normalfont In the evolutionary game with IPD\ with $T$ rounds as base game,
best response revision protocol, and $N$ players, the only absorbing states
are $\left(  N,0,0\right)  $, $\left(  0,N,0\right)  $ and $\left(
0,0,N\right)  $. Furthermore, if we define the following subsets of $S$%
\begin{align*}
S_{1}  &  =\{ \left(  s_{1},0,s_{3}\right)  :0\leq s_{1}\leq N,0\leq s_{3}\leq
N\},\\
S_{2}  &  =\left \{  \left(  s_{1},s_{2},s_{3}\right)  :2s_{1}+s_{2}%
<N+1\right \}  ,
\end{align*}
and assume $\left(  \mathbf{s}\left(  j\right)  \right)  _{j=1}^{\infty}$
starts at state $\mathbf{s}$, then:
\[
\text{if }\mathbf{s}\in S_{1}\cup S_{2}\text{ then }\Pr \left(  \mathbf{s}%
\left(  j\right)  \text{ is absorbed into either }\left(  N,0,0\right)  \text{
or }\left(  0,0,N\right)  \right)  =1.
\]

\end{proposition}

\subsection{Best Response Protocol, Influence of Payoff Matrix\label{sec0303}}

We now present the STG of an evolutionary game which is identical to the ones
of the previous section, except that we use $N=30$ players and the payoff
bimatrix
\[
\left(  A,A^{T}\right)  =\left(  \left[
\begin{array}
[c]{cc}%
a & 1\\
4 & 2
\end{array}
\right]  ,\left[
\begin{array}
[c]{cc}%
a & 4\\
1 & 2
\end{array}
\right]  \right)  .
\]
Here $a\in \{3.0,3.2,3.5,3.8\}$ is a game parameter.

\vspace{5mm}

\begin{minipage}{1.0\textwidth}
\centering
\includegraphics[scale=0.55]{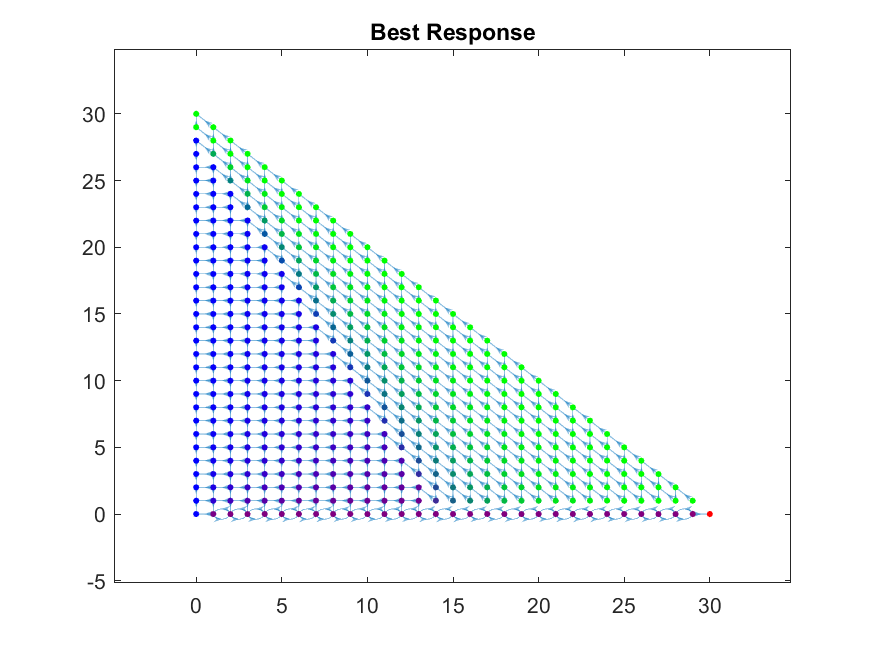}
\includegraphics[scale=0.55]{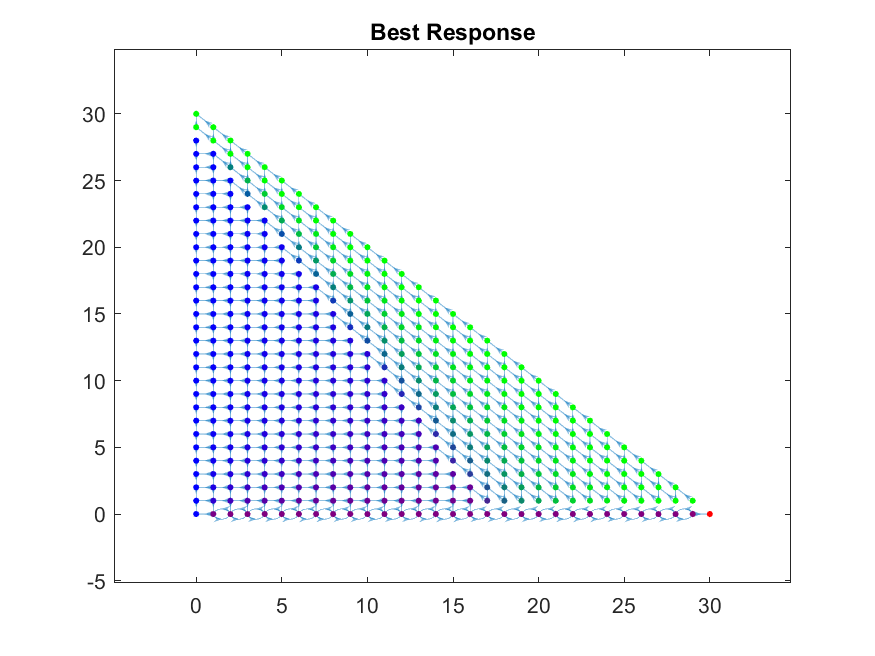}
\includegraphics[scale=0.55]{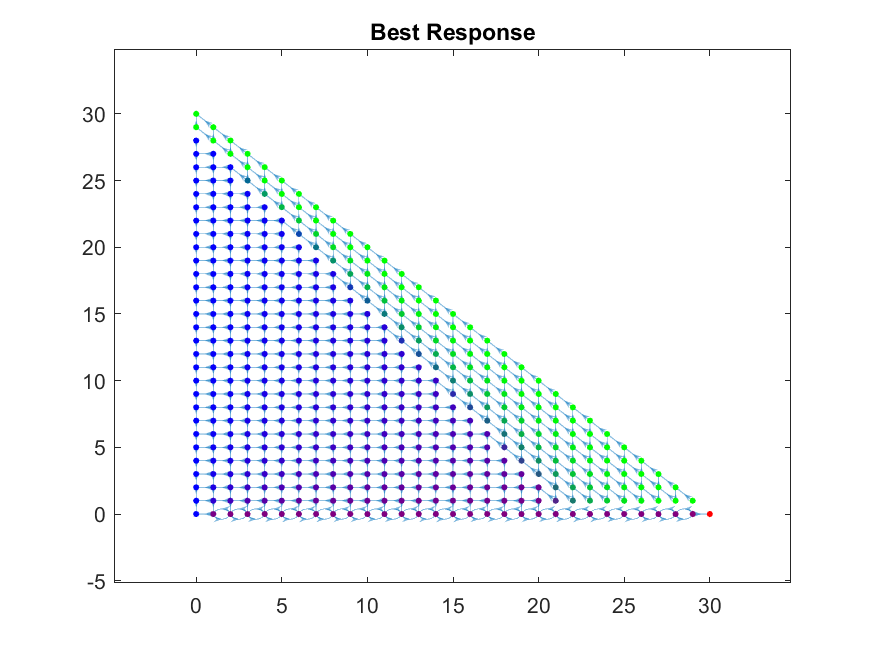}
\includegraphics[scale=0.55]{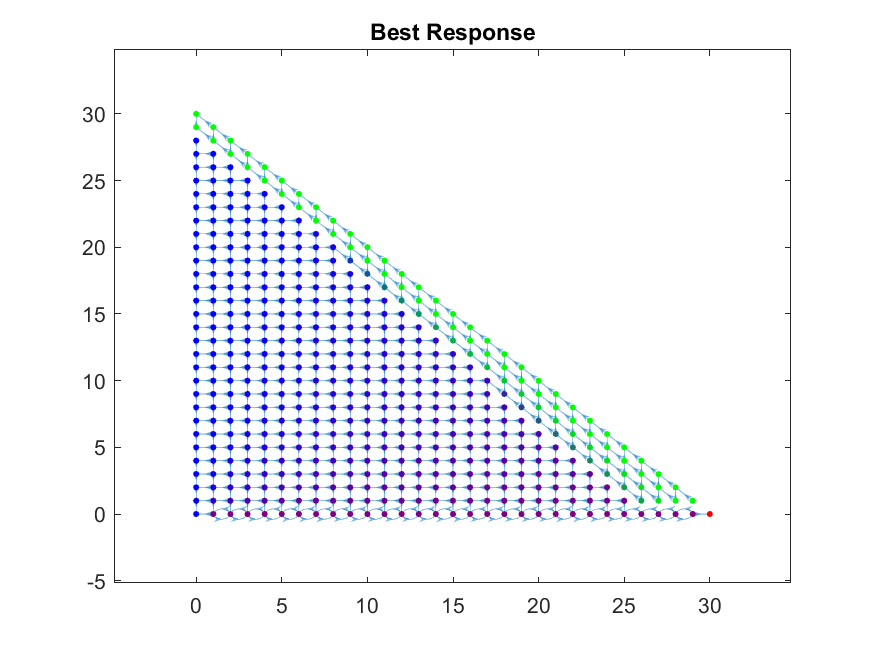}
\captionof{figure}{
STG diagrams for $N=30$ players and $a\in \{3.0,3.2,3.5,3.8\}$.
}
\label{fig004}%
\end{minipage}

\vspace{5mm}

\noindent Note the changein the basin of attraction of the pure AllD state
$\left(  0,3,0\right)  $:\ it gets smaller as $a$ increases. This is
reasonable, since a larger $a$ value encourages cooperation.

\subsection{Other Revision Protocols\label{sec0304}}

In this section we agan use the bimatrix
\[
\left(  A,A^{T}\right)  =\left(  \left[
\begin{array}
[c]{cc}%
3 & 1\\
4 & 2
\end{array}
\right]  ,\left[
\begin{array}
[c]{cc}%
3 & 4\\
1 & 2
\end{array}
\right]  \right)  .
\]
and we use $N=15$ players, but we experiment with various revision protocols.
In Figure \ref{fig005} we present the STG's corresponding to the revision
protocols PPC, PC, CAV, Logit. Compare the STG's to the one of Figure
\ref{fig003}.b, obtained from the BR\ protocol. We see that the change of the
revision protocol can result in a significant change of the STG\ structure
(consider especially the Logit STG in Figure \ref{fig005}.d). Also note the
existence of many more absorbing states (they are painted black).

\bigskip

\begin{minipage}{1.0\textwidth}
\centering
\includegraphics[scale=0.55]{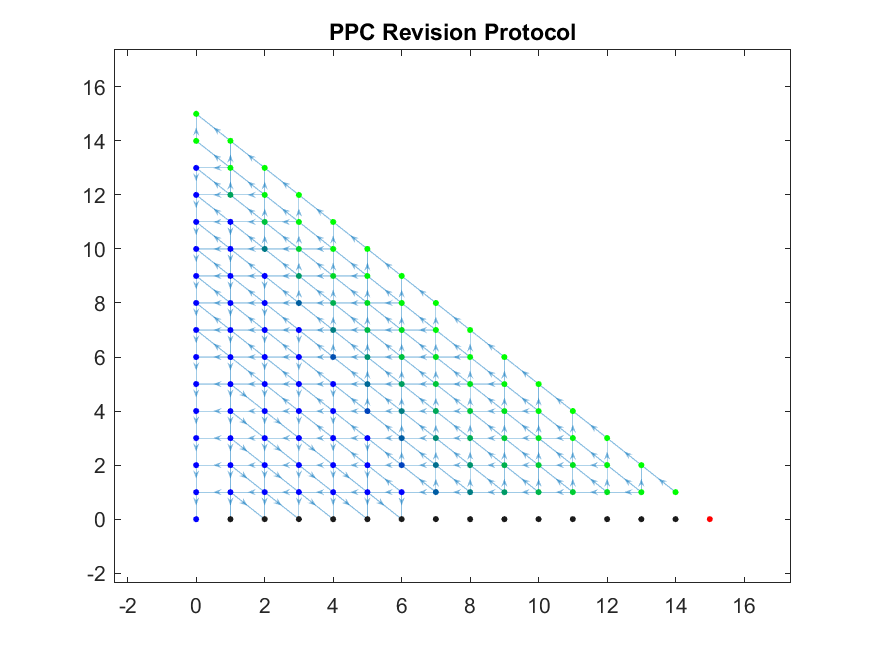}
\includegraphics[scale=0.55]{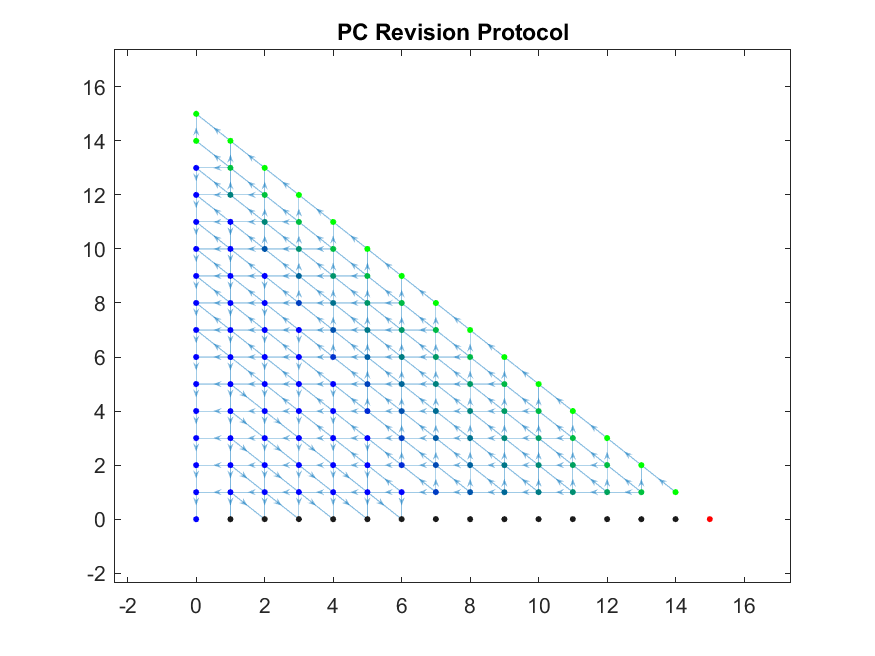}
\includegraphics[scale=0.55]{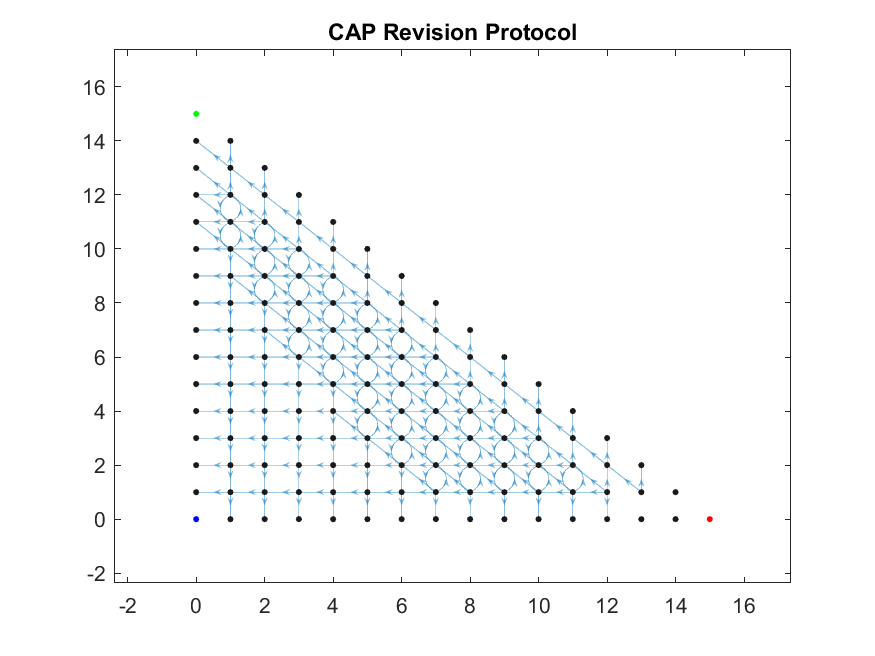}
\includegraphics[scale=0.55]{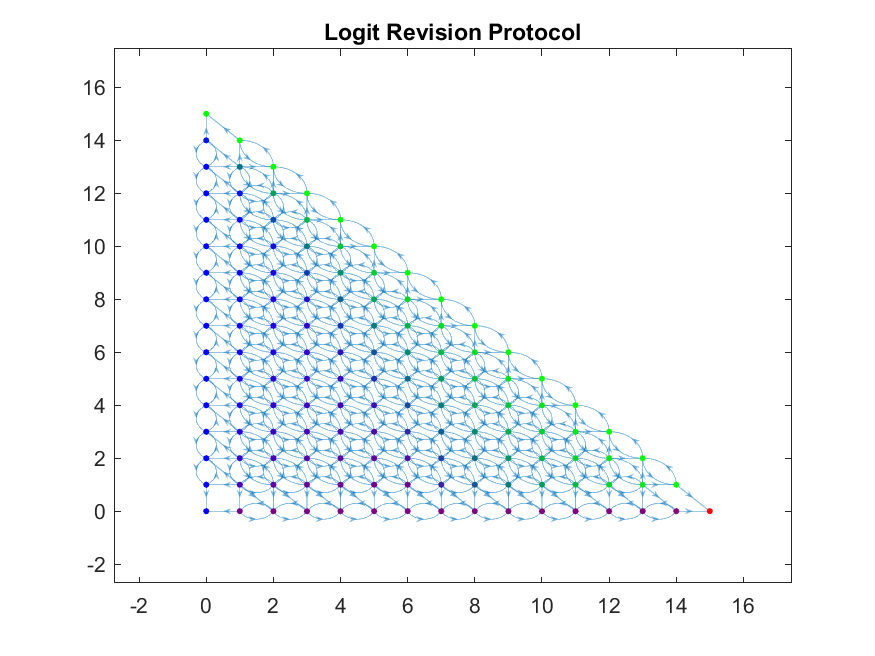}
\captionof{figure}{
STG diagrams for $N=15$ players and the revision protocols PPC, PC, CAV, Logit.
}
\label{fig005}%
\end{minipage}

\section{Stag Hunt\label{seg04}}

We now present an experiment in which the base game is the Stag Hunt with
payoff bimatrix
\[
\left(  A,A^{T}\right)  =\left(  \left[
\begin{array}
[c]{cc}%
10 & 1\\
8 & 5
\end{array}
\right]  ,\left[
\begin{array}
[c]{cc}%
10 & 8\\
1 & 5
\end{array}
\right]  \right)  .
\]
We set the strategy set to be \{AllC, AllD, TitForTat\} and always use $N=15$
players. We perform two experiments.

\begin{enumerate}
\item In the first experiment we use the above payoff bimatrix and experiment
with the revision protocols BR, PC, CAP, Logit. The corresponding results are
plotted in Figure \ref{fig006}.

\item In the second experiment we always use the BR\ revision protocol and use
the payoff bimatrix
\[
\left(  A,A^{T}\right)  =\left(  \left[
\begin{array}
[c]{cc}%
a & 1\\
8 & 5
\end{array}
\right]  ,\left[
\begin{array}
[c]{cc}%
a & 8\\
1 & 5
\end{array}
\right]  \right)
\]
with $a\in \left \{  8.5,9,10,11\right \}  $. The corresponding results are
plotted in Figure \ref{fig006}.
\end{enumerate}

\bigskip

\begin{minipage}{1.0\textwidth}
\centering
\includegraphics[scale=0.45]{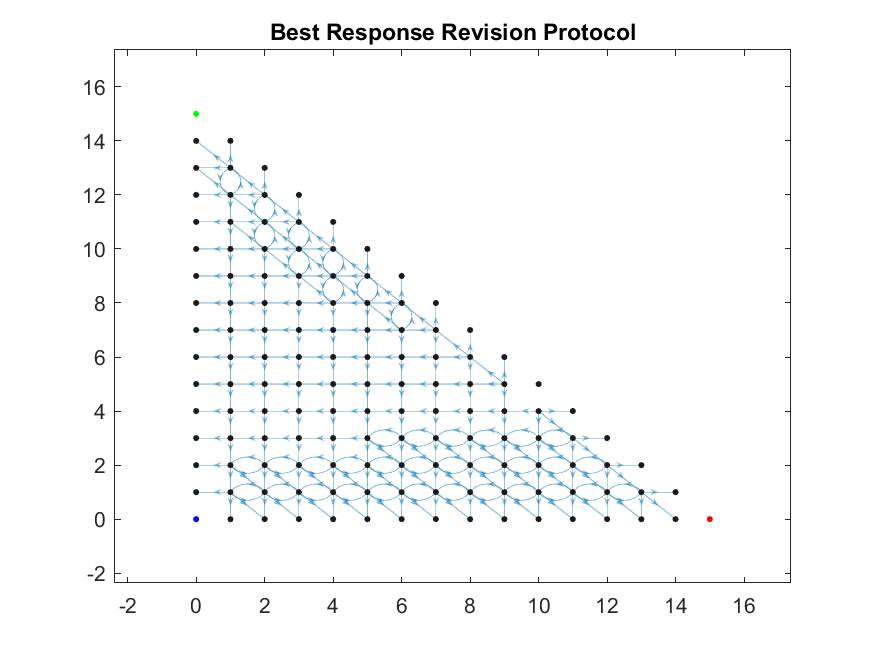}
\includegraphics[scale=0.45]{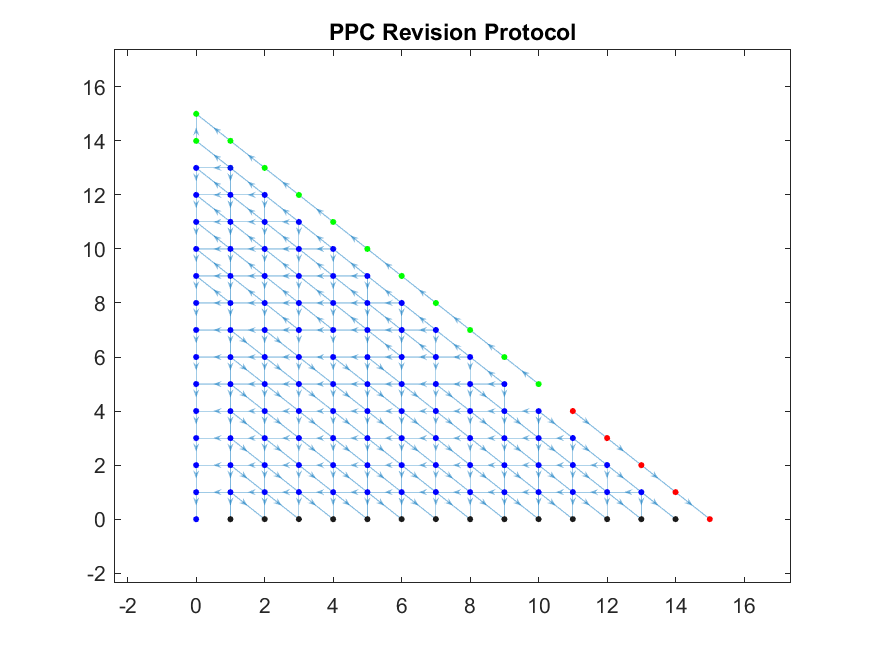}
\includegraphics[scale=0.45]{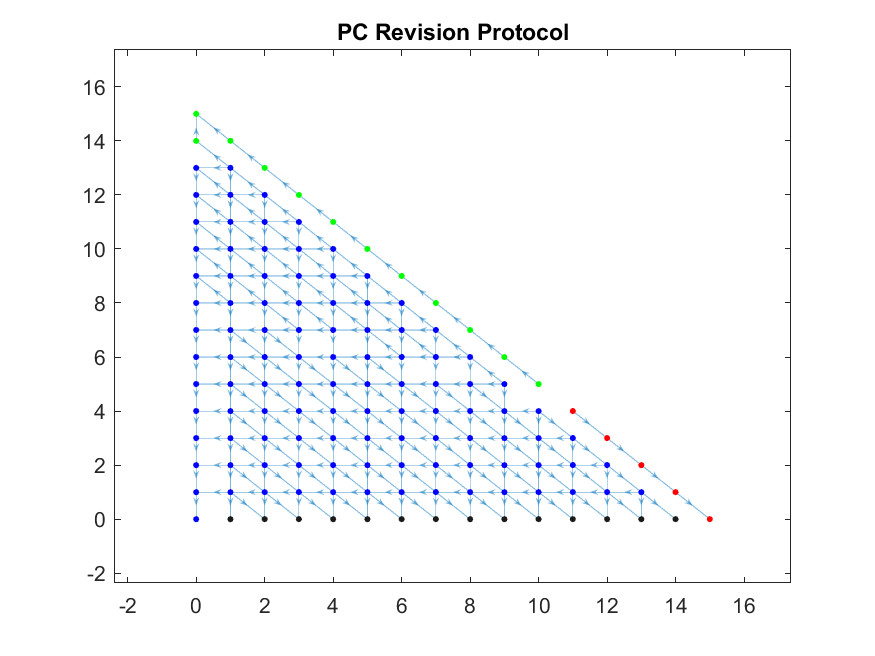}
\includegraphics[scale=0.45]{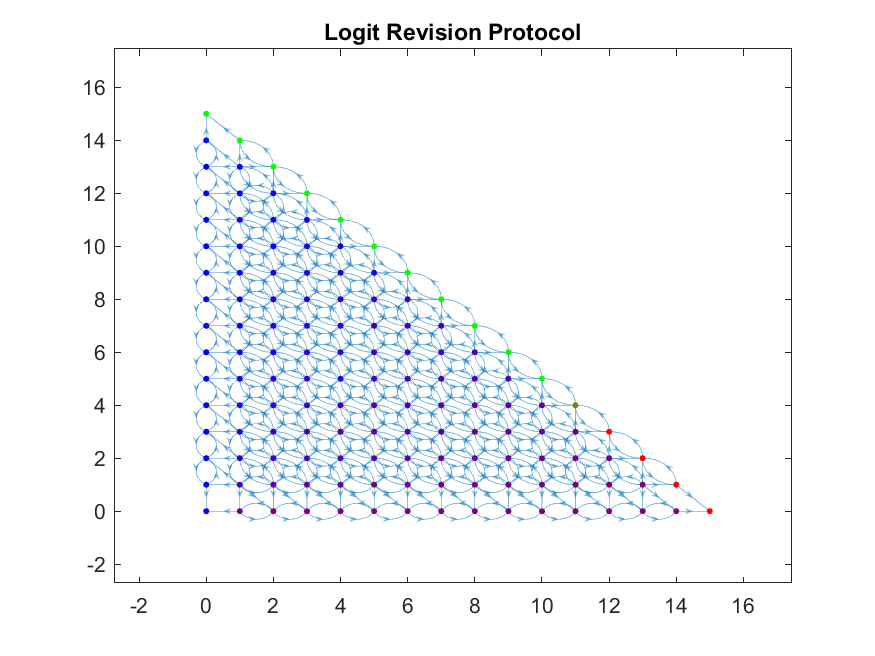}
\captionof{figure}{
STG diagrams for Stag Hunt with $N=15$ players and the revision protocols BR, PPC, PC, Logit.
}
\label{fig007}%
\end{minipage}

\bigskip

\begin{minipage}{1.0\textwidth}
\centering
\includegraphics[scale=0.45]{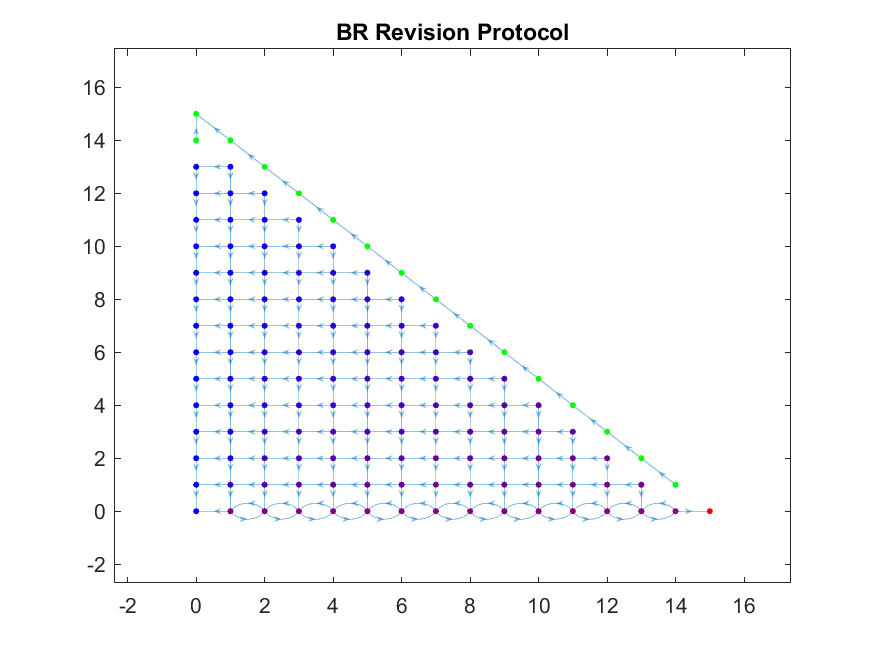}
\includegraphics[scale=0.45]{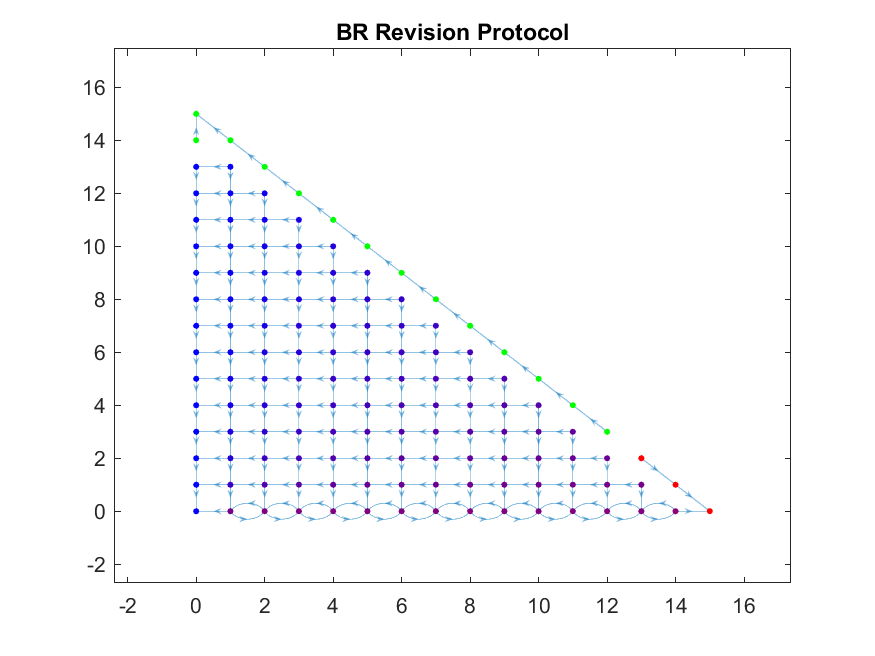}
\includegraphics[scale=0.45]{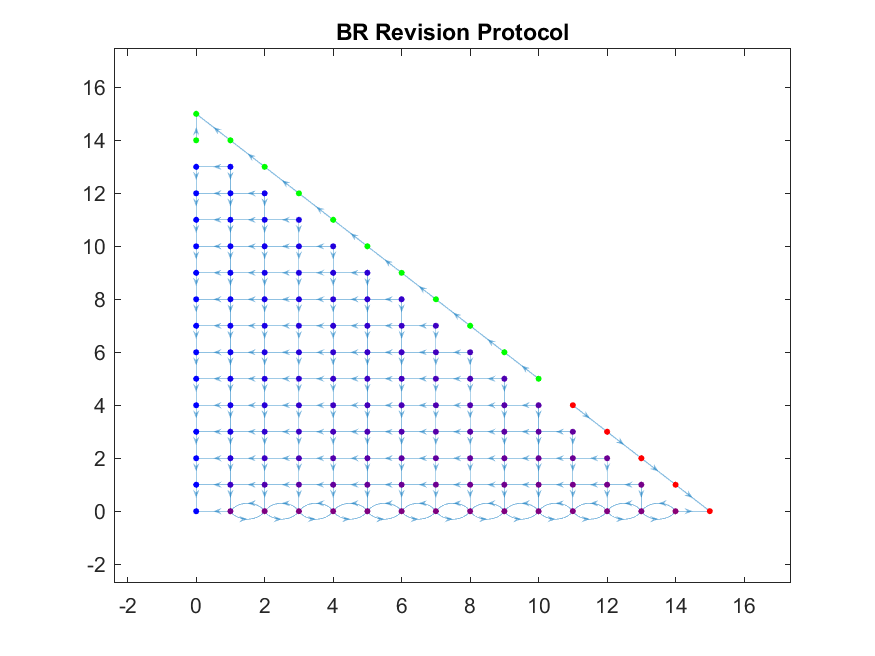}
\includegraphics[scale=0.45]{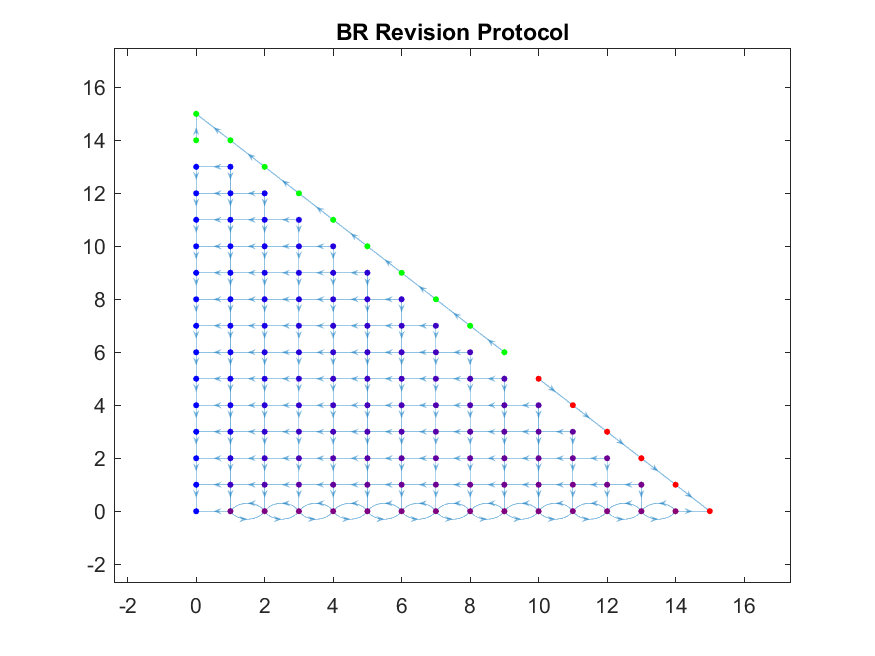}
\captionof{figure}{
STG diagrams for Stag Hunt with $N=15$ players, the revision protocols BR and varying payoff matrix $A$.
}
\label{fig007}%
\end{minipage}

\vspace{5mm}

\section{Rock-Paper-Scissors\label{sec05}}

In the previous examples we have studied evolutionary games in which the $B$
matrix is obtained from an iterated game (iterated Prisoner's Dilemma,
iterated Stag Hunt). In this example we start with a one-round game, namely
the Rock-Paper-Scissor game which has payoff bimatrix
\[
\left(  B,-B\right)  =\left(  \left[
\begin{array}
[c]{rrr}%
0 & -1 & 1\\
1 & 0 & -1\\
-1 & 1 & 0
\end{array}
\right]  ,\left[
\begin{array}
[c]{rrr}%
0 & 1 & -1\\
-1 & 0 & 1\\
1 & -1 & 0
\end{array}
\right]  \right)  .
\]
There are three pure strategies available to both players, labeled as Rock,
Paper and Scissors.

The transition probability matrix $P$ corresponding to Rock-Paper-Scissors has
an interesting structure, which is easiest to understand by looking at the
STG's portrayed in Figure \ref{fig008}; these are the STG's obtained from the
Best Response protocol, for $N\in \{3,4,5,6,8,15,60,75\}$ players. Note the
symmetric nature of the STG for all $N$ values. Also note that, for all $N$,
the only absorbing states are the pure ones; this implies that in every
realization the entire population will eventually adopt a single strategy.

\vspace{5mm}

\begin{minipage}{1.0\textwidth}
\centering
\includegraphics[scale=0.45]{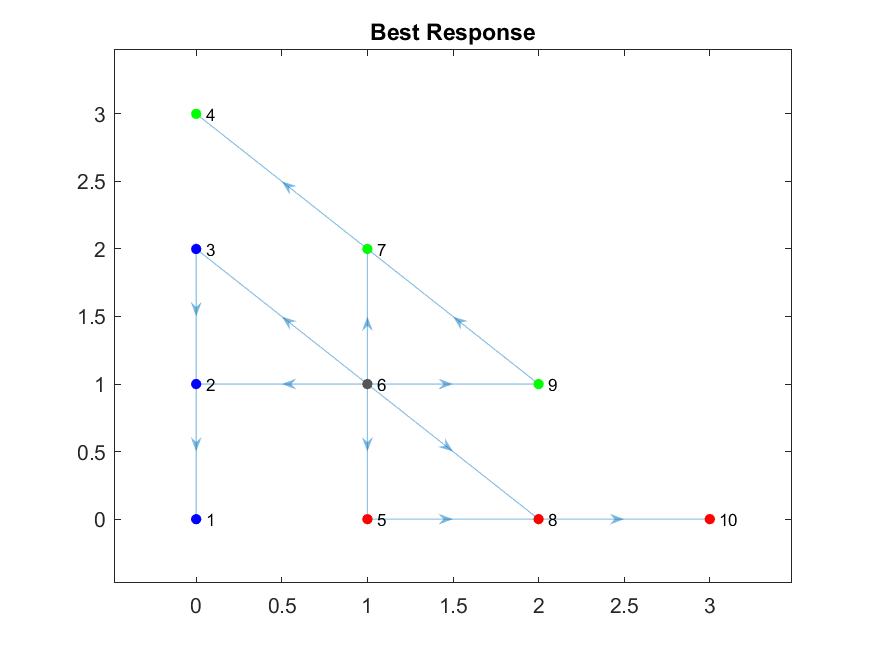}
\includegraphics[scale=0.45]{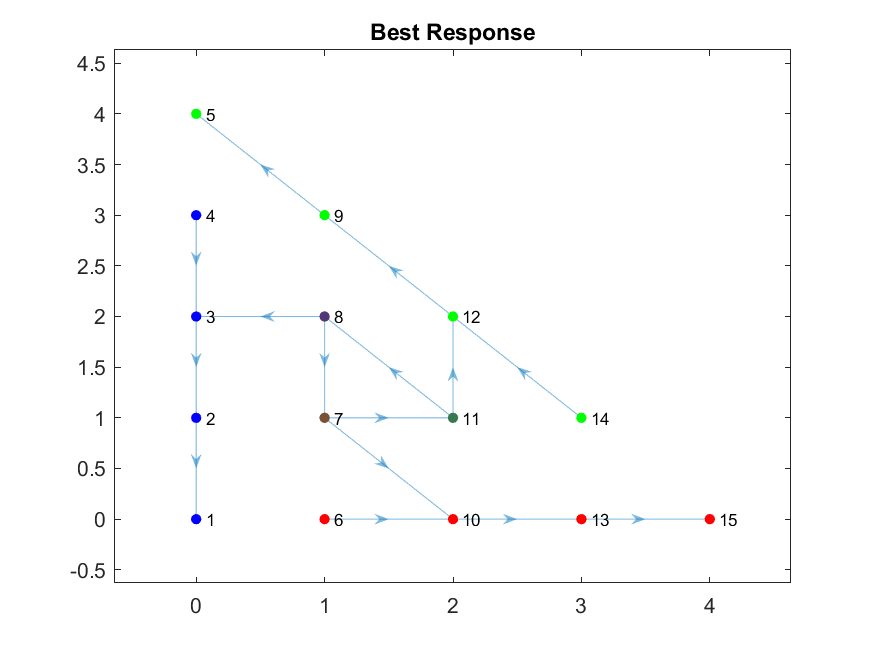}
\end{minipage}

\begin{minipage}{1.0\textwidth}
\centering
\includegraphics[scale=0.45]{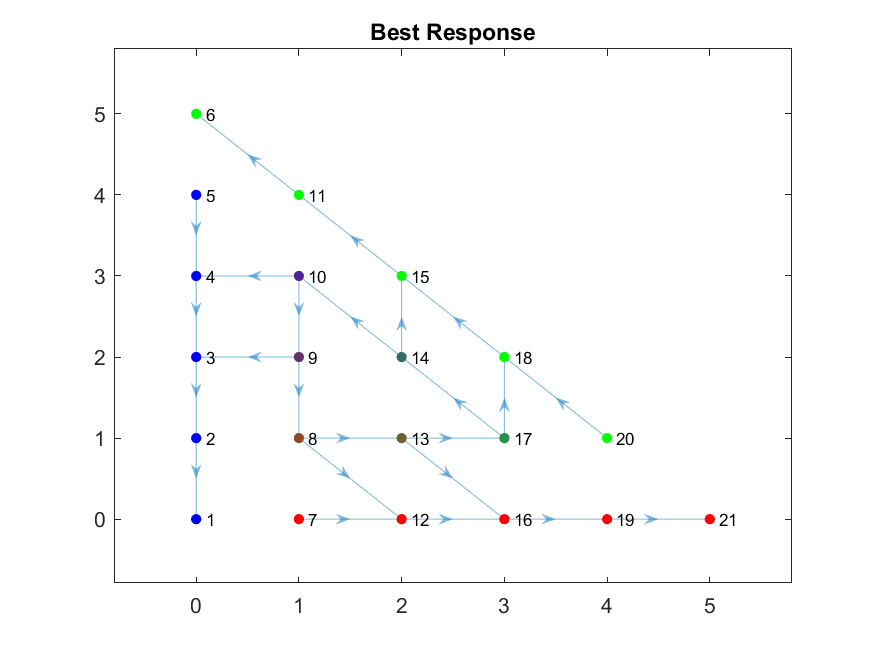}
\includegraphics[scale=0.45]{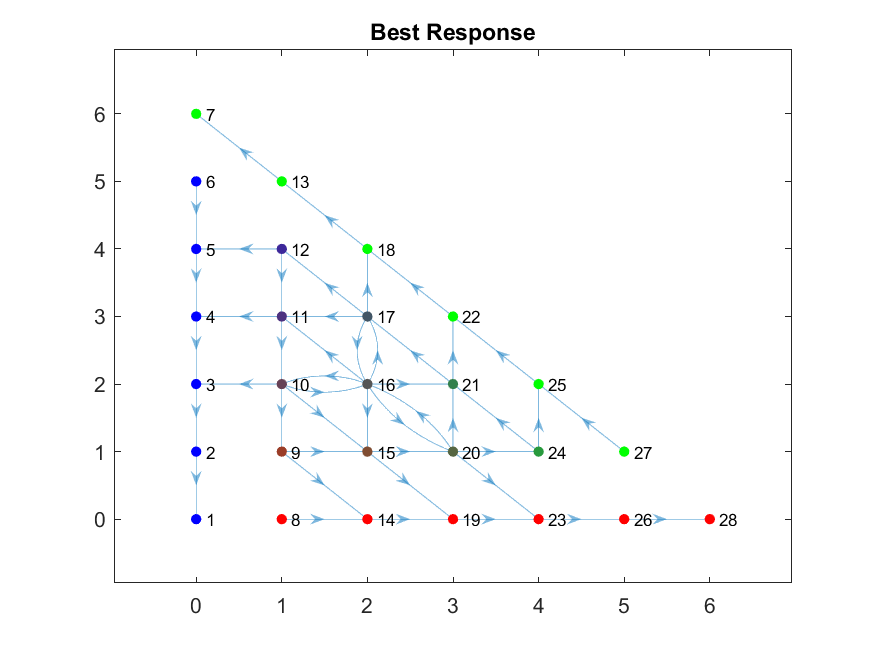}
\end{minipage}

\begin{minipage}{1.0\textwidth}
\centering
\includegraphics[scale=0.45]{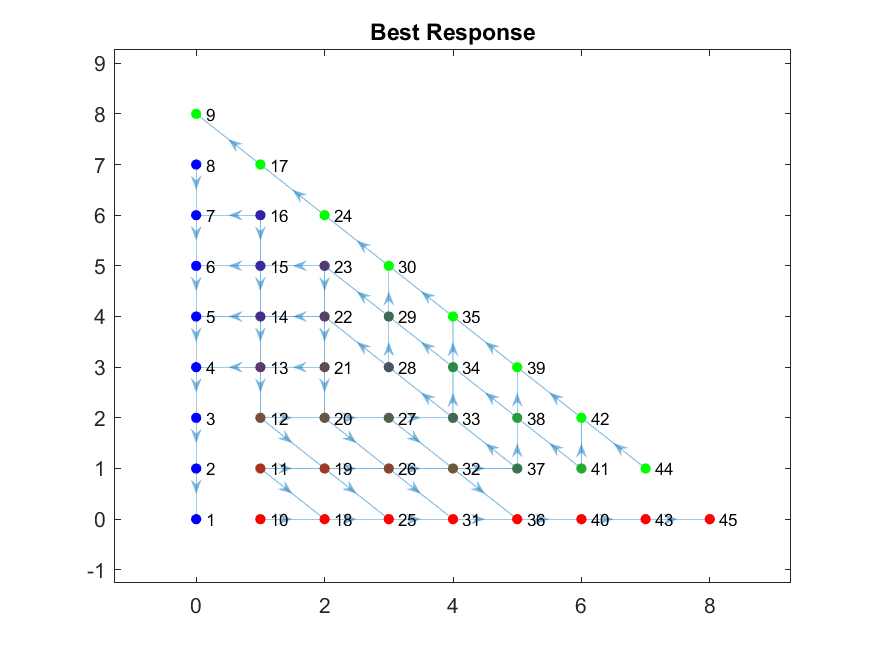}
\includegraphics[scale=0.45]{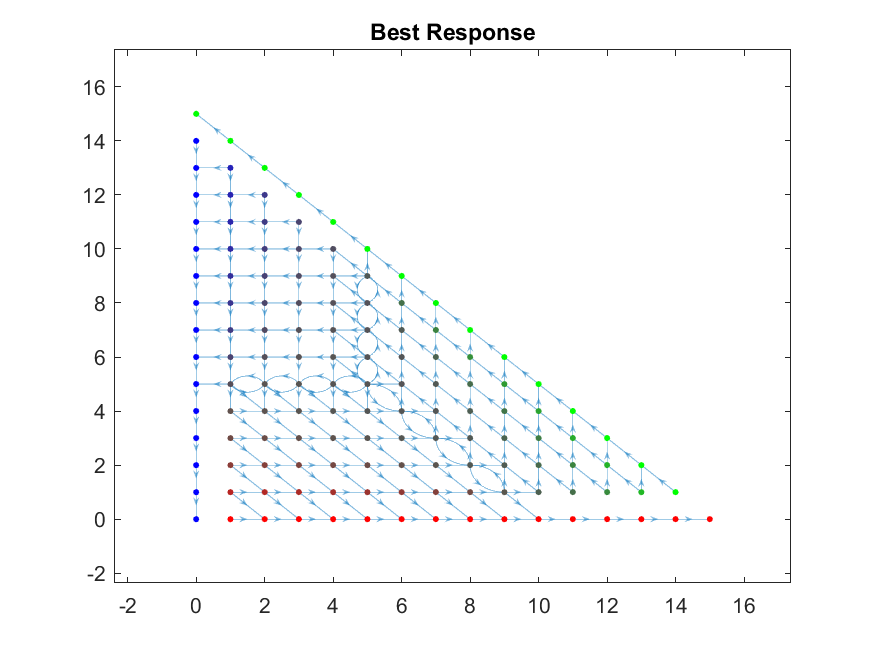}
\end{minipage}

\begin{minipage}{1.0\textwidth}
\centering
\includegraphics[scale=0.45]{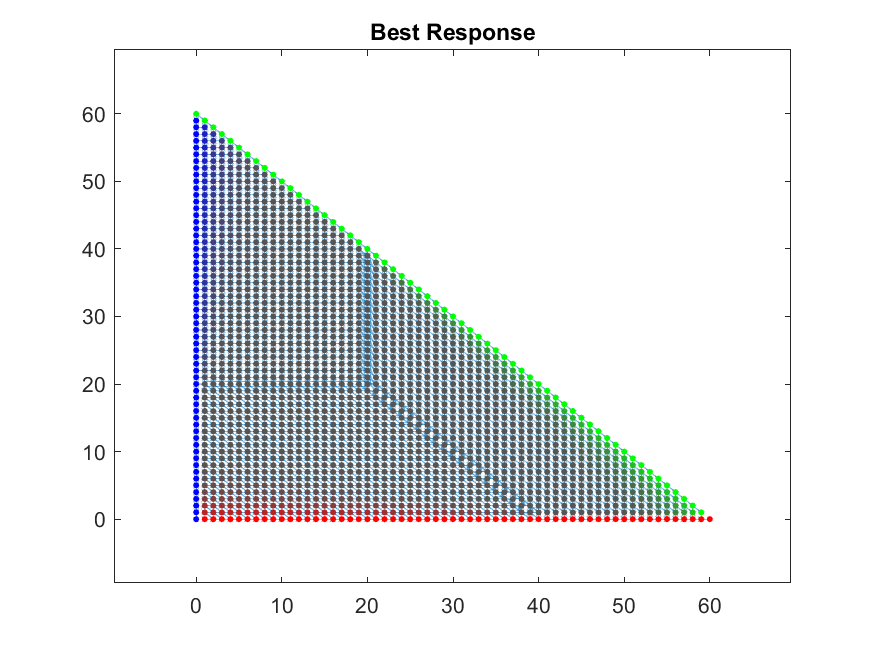}
\includegraphics[scale=0.45]{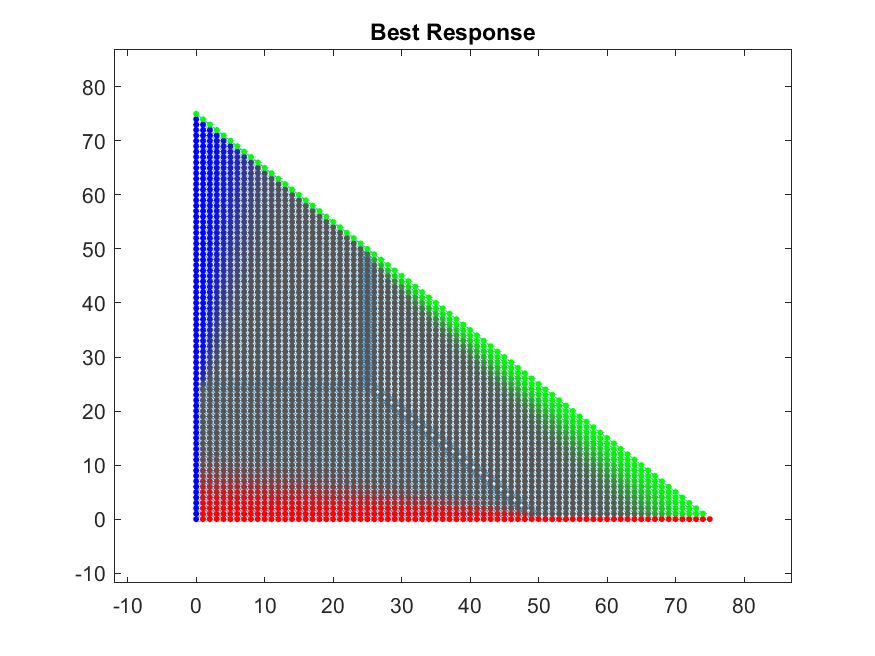}
\captionof{figure}{
STG's of the evolutionary game with Rock-Paper-Scissors as base game, Best Response protocol and
for $N\in \{3,4,5,6,8,15,60,75\}$ players.
}
\label{fig008}
\end{minipage}


\begin{thebibliography}{9}                                                                                                %


\bibitem {Axelrod1981}R. Axelrod and W. D. Hamilton. \textquotedblleft The
evolution of cooperation.\textquotedblright \  \emph{Science}, vol. 211, pp.
1390-1396, 1981.

\bibitem {Hofbauer1998}J. Hofbauer and K. Sigmund. \emph{Evolutionary games
and population dynamics}. Cambridge University Press, 1998.

\bibitem {Kemeny1966}J.G. Kemeny and J. L. Snell. \emph{Finite markov chains}.
Princeton, NJ: van Nostrand, 1969.

\bibitem {Sandholm2010a}W.H. Sandholm, \emph{Population Games and Evolutionary
Dynamics}. MIT\ Press, 2010.

\bibitem {Sandholm2010b}W.H. Sandholm. \textquotedblleft Local stability under
evolutionary game dynamics\textquotedblright.\  \emph{Theoretical Economics},
vol. 5, pp. 27-50, 2010.
\end{thebibliography}
\end{document}